\documentclass{article}

\usepackage{hyperref}
\usepackage[authoryear]{natbib}
\usepackage[shortlabels]{enumitem}
\usepackage{amssymb}
\usepackage{float}
\usepackage{graphicx} 
\usepackage[multidot]{grffile}
\usepackage{authblk}
\usepackage[margin=0.8in]{geometry}
\usepackage{indentfirst}
\usepackage{chemformula} % Writing Molecule Names with Subscripts and Superscripts, i.e. rather than writing H$_2$O we write \ch{H2O} with this package

\DeclareSymbolFont{matha}{OML}{txmi}{m}{it}% txfonts
\DeclareMathSymbol{\varv}{\mathord}{matha}{118}

\title{\textbf{A HITRAN-formatted UV line list of \ch{S2} containing transitions involving $X\,^{3}\Sigma^{-}_{g}$, $B\,^{3}\Sigma^{-}_{u}$, and $B''\,^{3}\Pi_{u}$ electronic states}}

\author[1,2]{Frances M. Gomez}
\author[1]{Robert J. Hargreaves}
\author[1,*]{Iouli E. Gordon}
\affil[1]{Atomic and Molecular Physics, Center for Astrophysics $\vert$ Harvard \& Smithsonian, Cambridge, MA USA}
\affil[2]{University of New Hampshire, Chemistry Department, Durham, NH 03824, USA}
\affil[*]{Corresponding author E-mail address: igordon@cfa.harvard.edu}

\date{\vspace{-4mm}\small\today}

\begin{document}
\maketitle
\vspace{-6mm}

%%%%%%%Abstract%%%%%%
\renewenvironment{abstract}
 {\small
  \begin{center}
  \bfseries \abstractname\vspace{-.5em}\vspace{0pt}
  \end{center}
  \list{}{%
    \setlength{\leftmargin}{0.5mm}
    \setlength{\rightmargin}{\leftmargin}
}
  \item\relax}
 {\endlist}
 \vspace{4mm}
\begin{abstract}
\centering
The sulfur dimer (\ch{S2}) is an important molecular constituent in cometary atmospheres and volcanic plumes on Jupiter's moon Io. It is also expected to play an important role in the photochemistry of exoplanets. The UV spectrum of \ch{S2} contains transitions between vibronic levels above and below the dissociation limit, giving rise to a distinctive spectral signature.  By using spectroscopic information from the literature, and the spectral simulation program PGOPHER, a UV line list of \ch{S2} is provided. This line list includes the primary $B\,^{3}\Sigma^{-}_{u}-X\,^{3}\Sigma^{-}_{g}$ ($v'$=0-27, $v''$=0-10) electronic transition, where vibrational bands with  $v'$$\geq$10 are predissociated. Intensities have been calculated from existing experimental and theoretical oscillator strengths, and semi-empirical strengths for the predissociated bands of \ch{S2} have been derived from comparisons with experimental cross-sections. The \ch{S2} line list also includes the $B''\,^{3}\Pi_{u}-X\,^{3}\Sigma^{-}_{g}$ ($v'$=0-19, $v''$=0-10) vibronic bands due to the strong interaction with the $B$ state. In summary, we present the new HITRAN-formatted \ch{S2} line list and its validation against existing laboratory spectra. The extensive line list covers the spectral range 21\,700$-$41\,300~cm$^{-1}$ ($\sim$242$-$461~nm) and can be used for modeling both absorption and emission.
\end{abstract}

%%%%%%%%%% Introduction %%%%%%%%
\section*{\small Introduction}\vspace{-1.5mm}\hrule\vspace{3mm}\label{sec:intro}
The sulfur dimer, disulfur (\ch{S2}), has been observed in ultraviolet (UV) spectra of comets such as IRAS-Araki-Alcock 1983d \citep{10.1086/184158} and Hyakutake \citep{10.1029/98GL01953}. The impact of comet Shoemaker-Levy 9 with Jupiter produced large amounts of \ch{S2} in the stratosphere \citep{10.1126/science.7871428} leading to a rich sulfur chemistry \citep{1995GeoRL..22.1593Z}. In addition, \ch{S2} has been detected in volcanic plumes on Jupiter’s moon Io \citep{10.1126/science.288.5469.1208}.

\ch{S2} is a key intermediary in the exothermic polymerization of elemental sulfur en route to octasulfur, \ch{S8} \citep{10.1007/BF01808144,Shingledecker2020}, the stable molecular form. The polymerization of sulfur towards \ch{S8} can be interrupted by the photolysis of \ch{S2}, and it has also been noted that S$_n$O can photolyze to S$_n$ \citep{10.1038/s41467-022-32170-x}. Although, formation pathways of polysulfur molecules (including \ch{S2}) in the atmosphere of Venus are still under debate \citep{10.1038/s41467-022-32170-x}, it has been suggested that photodissociation of \ch{S2} could fill a needed gap in Venusian atmospheric models \citep{10.1038/s41467-022-32170-x}. While \ch{S2} may only be an intermediary observation, its UV absorption can help infer sulfur chemistry even without the detection of \ch{S8}.

\citet{Hobbs2021} have investigated thermochemical and photochemical sulfur reactions in the atmospheres of warm and hot Jupiters. They found that at 10$^{-3}$ bar and at temperatures around 1000 K, mixing ratios of \ch{S2} can be up to 10$^{-5}$.  The recent detection of \ch{CO2} \citep{2023Natur.614..649J}, and apparent \ch{SO2} absorption feature \citep{2022arXiv221110487R}, in the atmosphere of exoplanet WASP-39b has increased the need to better understand the spectroscopy of photochemically produced sulfur species. In particular, for WASP-39b it has been predicted that \ch{S2} is a key molecule in the photochemical pathway to forming \ch{SO2} and is expected to be the most abundant sulfur-containing molecule at pressures probed by JWST transmission spectra during the evening terminator \citep{2022arXiv221110490T}. 

While many works have investigated the energy levels and spectrum of \ch{S2}, including the analysis of perturbations \citep[e.g.,][]{10.1063/1.470810} and predissociated bands \citep[e.g.,][]{10.1063/1.5029930}, an accurate \ch{S2} line list that is capable of reproducing the UV spectrum of \ch{S2} at high resolution is currently unavailable in the literature or public databases. This was highlighted by \citet{Kim2003} when building their fluorescence model for analyses of cometary spectra. \citet{Kim2003} extended their atlas \citep{Kim1994} by including limited experimental information from the literature, which included some direct entries from early experimental works \citep{Ikenoue1953,Ikenoue1960}. However, as acknowledged by the authors, their model was still very limited, particularly in terms of accounting for perturbations. Similarly, in analyses of Io spectra, \citet{10.1126/science.288.5469.1208} used an unpublished line list that was calculated for the purposes of their work. Perturbations were not accounted for and some of the experimental intensity information which now exists was not available at that time \citep[e.g.,][]{10.1063/1.5029929}.  Recently, \citet{sarka_2023} have released an \textit{ab initio} line list of \ch{S2}, but spin-spitting and perturbation effects have not been accounted for, which limits the accuracy for high resolution applications.  

\ch{S2} photolysis has previously been estimated in photochemical models based on inferences from solar system cometary analyses \citep{10.1007/BF00057603,10.1073/pnas.0903518106,10.1088/0004-637X/761/2/166} and have been scaled by the actinic flux at 300~nm. Use of this approximation can lead to inaccurate photolysis rates on planets orbiting stars of other spectral types (e.g., M-dwarfs), where the shape of the spectral energy distribution (SED) is different than the Sun. Another consequence is that it disregards the photochemical self-shielding, which is caused by overlying \ch{S2} and/or other overlying molecules. \citet{Hobbs2021} employed the calculated cross-sections from the Leiden database \citep{Heays2017}, but it should be noted that the photodissociation cross-sections measured subsequently in \citet{10.1063/1.5029929} differ substantially from those calculated in \citet{Heays2017}.  

The goal of this work is to provide a reliable publicly available line list of \ch{S2}. A line-by-line parameterization, such as that employed by the HITRAN database \citep{10.1016/j.jqsrt.2021.107949}, has advantages over available parametrizations for \ch{S2} spectra. These line lists are compatible with a majority of community radiative transfer codes, thereby allowing the generation of cross-sections at a variety of thermodynamic conditions and over a wide spectral range. One of the peculiarities of parameterizing the UV line list of the sulfur dimer is that it has to be capable of simulating the UV spectrum at high resolution below the dissociation limit, while also reproducing the diffuse spectrum above the dissociation limit where rovibronic features can not be resolved.  Indeed, the predissociation widths of the disulfur transitions that have upper levels above the dissociation limit have widths that reach tens of wavenumber, therefore obscuring the resolved structure of individual transitions. The \ch{S2} line list from this work will, therefore, need to be capable of reproducing these effects. 

%%%%%%%%%%%%%%%%%%%%%%%%%%%%%%%%%%%%%%%%%%%%%%%%%%

\subsection*{\small Spectroscopy of the \texorpdfstring{\ch{S2}}{S2} molecule}\vspace{-1.5mm}\hrule\vspace{3mm} \label{sec:modiagrams}
\ch{S2} is isoelectronic to molecular oxygen, \ch{O2}. The two molecules have similar electronic states and exhibit similar visible and UV bands. Thus, comparisons between \ch{S2} and the much better studied \ch{O2} system provide valuable insights when generating a comprehensive \ch{S2} line list. Unlike oxygen, \ch{S2} is an unstable molecule. In the laboratories it is produced in sulfur-containing flames and discharges since it mainly forms at high temperatures (800~K) \citep{10.1063/1.476074}. The $B-X$ transition is quite intense and emits a blue color as seen in experimental and cometary spectra. 

\begin{figure}[ht]
    \centering
	\includegraphics[width=8cm]{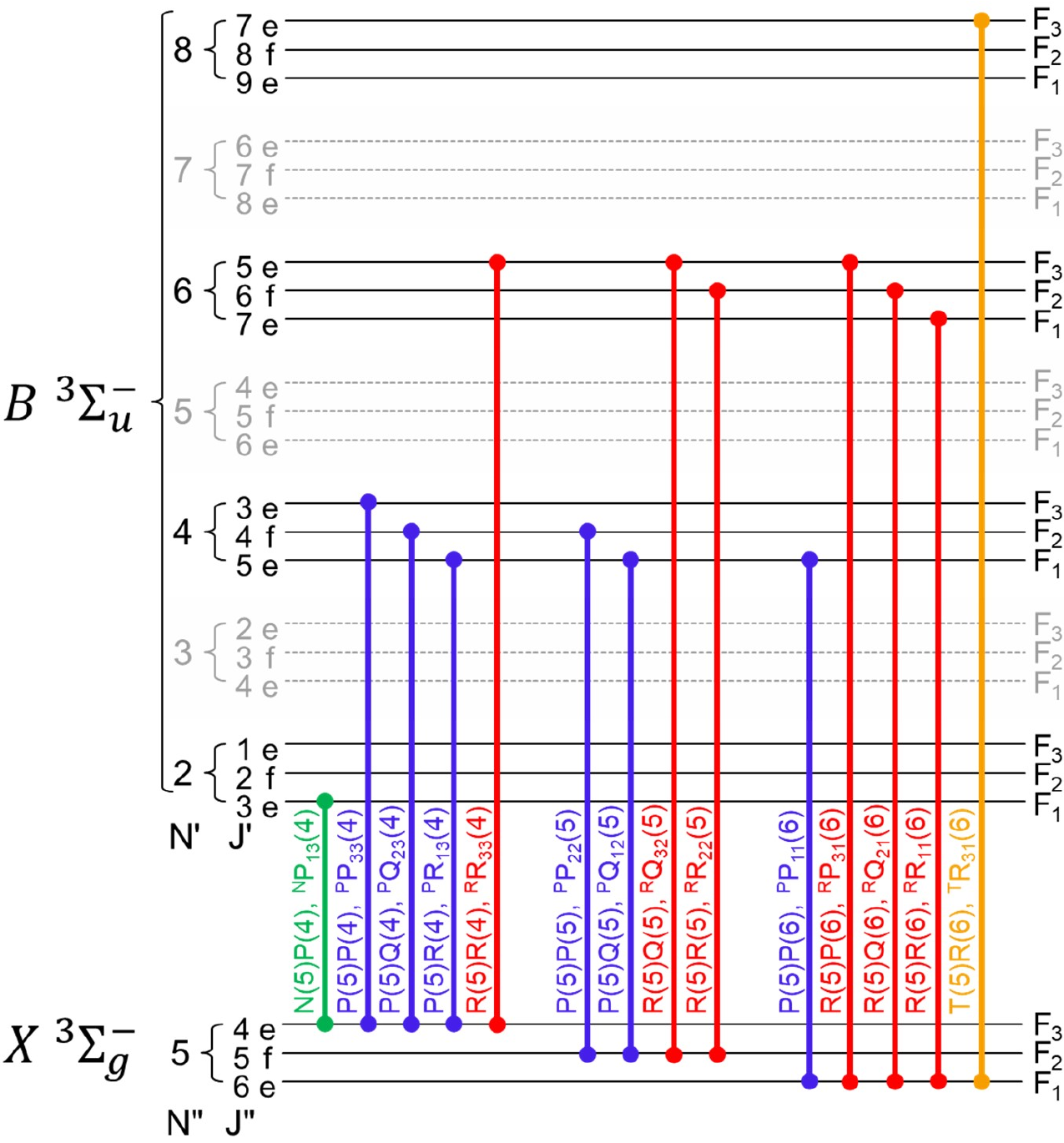}
    \caption{A schematic showing the possible rovibronic transitions for the $B\,^{3}\Sigma^{-}_{u} - X\,^{3}\Sigma^{-}_{g}$ transition of \ch{S2} with $N''=5$. The $^{{\Delta}N}{\Delta}J_{F'F''}(J'')$ and $\Delta N$($N''$)$\Delta J$($J''$) notations are both shown for each transition, with the transition color relating to $\Delta N$. Even $N$ levels of the $X\,^{3}\Sigma^{-}_{g}$ state and odd $N$ levels of the $B\,^{3}\Sigma^{-}_{u}$ (greyed and dashed lines) are not populated. Note that the depicted relative positions of spin components of different rotational levels, are qualitative and do not represent the actual relative energies, especially in the perturbed upper electronic state. }
    \label{fig:levels}
\end{figure}

Figure 1 of \citet{10.1021/acs.jpca.9b05350} provides a detailed comparative overview of the lowest electronic states of \ch{S2} and O$_{2}$ molecules. Considering that in this work, we concentrate only on the part of the UV spectrum, here, we provide only a brief summary of the states of \ch{S2}. Just as in the case of molecular oxygen, the ground electronic state of disulfur is $X\,^{3}\Sigma^{-}_{g}$. The lowest excited states are two singlet states $a\,^{1}\Delta_{g}$ and $b\,^{1}\Sigma^{+}_{g}$ with $T_{e} = 4322.99$ and 7788.72~cm$^{-1}$, respectively \citep{2013MolPh.111..673X}. The electric dipole transitions involving these states and the ground state are spin-forbidden, however, much weaker transitions are possible through magnetic dipole and electric quadrupole mechanisms \citep{Setzer2003,10.1016/0022-2852(86)90032-9}.  Above them lie the $A\,^{3}\Sigma^{+}_{u}$, $A'\,^{3}\Delta_{u}$, and $c\,^{1}\Sigma^{-}_{u}$ states, which, in the oxygen molecule, are responsible for the so-called Herzberg bands. The electronic state of our interest is $B\,^{3}\Sigma^{-}_{u}$ with term energy $T_{e} = 31967.36$~cm$^{-1}$, and transition between this state and the ground state are analogous to the so-called Schumann-Runge bands in molecular oxygen.  This state is in the close proximity to $B'\,^{3}\Pi_{g}$ and $B''\,^{3}\Pi_{u}$  \citep{10.1016/j.jqsrt.2019.106805} states and the interference from the latter has to be taken into account in our line list. At higher energies, near 41\,000~cm$^{-1}$, lie the $e\,^{1}\Pi_{g}$ and $f\,^{1}\Delta_{u}$ states.  S$_{2}$ also has a number of unbound electronic states that cross the bound $B\,^{3}\Sigma^{-}_{u}$ state at increasing energy in the order $1\,^{1}\Pi_{u}$,  $1\,^{3}\Pi_{g}$, $1\,^{1}\Pi_{g}$, $1\,^{5}\Sigma^{-}_{g}$, $1\,^{5}\Pi_{u}$, and $2\,^{3}\Sigma^{+}_{u}$ \citep{10.1021/acs.jpca.9b05350,10.1063/1.476074}.

In the HITRAN database, the quantum notation of transitions between states obeying Hund's case (b) is assumed for molecules in triplet states \citep[see Appendix of][]{10.1016/j.jqsrt.2021.107949}. Under this formulation, each rotational level \emph{N} is split into three spin components with total angular momentum \textbf{J=N+S}, and this notation is adopted here for consistency. In case (b) formalism, the spin components are defined as $F_1$: $J=N+1$, $F_2$: $J=N$, and $F_3$: $J=N-1$. However, as will be seen in the next section, unlike the case of \ch{O2}, the spin-spin coupling constant of \ch{S2} is much larger than the rotational constants causing the splitting of spin components of the individual rotational levels to be larger than the separation between these levels, therefore Hunds case (a) is more appropriate, at least for relatively low rotational levels. Indeed, although S$_{2}$ has a $\Sigma$ ground state, the total electron spin is equal to 1, therefore the projection of total angular momentum on the internuclear axis, $\Omega=\Lambda+\Sigma$ (where $\Lambda$ is a projection of orbital angular momentum \textbf{L}, and $\Sigma$ is the projection of total electron spin \textbf{S}), can be 0 or 1. Therefore, \ch{S2} can exhibit case (a) behavior. In case (a) formalism of intermediate coupling $F_{1,3}$ are a mixture of $\Omega=0$ and $\Omega=1$ components, whereas $F_2$ is pure $\Omega=1$. For the case when the spin components can be considered uncoupled (Hund's case (c)) $F_{1}$ corresponds to $\Omega=0$ only, whereas $F_{2,3}$ to $\Omega=1$. Just as for the \ch{^{16}O2} isotopologue in the ground electronic state, in \ch{^{32}S2} only levels with a total parity (+) are allowed due to nuclear spin statistics (nuclear spin of the $^{32}$S is zero), therefore, in the Hund's case (b) formalism, every other rotational level (i.e., the even levels) are missing. In the excited $B\,^{3}\Sigma^{-}_{u}$ state, total parity (+) levels will correspond to only even rotational states being populated, and the odd levels are missing. 

Selection rules require $\Delta$$J$ to be equal to 0 or $\pm$1 and 14 branches are possible for the $B\,^{3}\Sigma^{-}_{u} - X\,^{3}\Sigma^{-}_{g}$. Traditionally, in spectroscopic papers, the $^{{\Delta}N}{\Delta}J_{F'F''}(J'')$ notation for line assignments is used. In this notation, it is common to refer to six major branches between the same spin-components ($^{\textrm{R}}$R$_{11}$,
$^{\textrm{R}}$R$_{22}$,
$^{\textrm{R}}$R$_{33}$,
$^{\textrm{P}}$P$_{11}$,
$^{\textrm{P}}$P$_{22}$, and $^{\textrm{P}}$P$_{33}$) with eight weaker satellite branches ($^{\textrm{N}}$P$_{13}$,
$^{\textrm{R}}$P$_{31}$,
$^{\textrm{P}}$Q$_{12}$,
$^{\textrm{P}}$Q$_{23}$,
$^{\textrm{R}}$Q$_{32}$,
$^{\textrm{R}}$Q$_{21}$,
$^{\textrm{P}}$R$_{13}$, and $^{\textrm{T}}$R$_{31}$).

However, in our line list, we employ the $\Delta N$($N''$)$\Delta J$($J''$) notation, which is closer in appearance to quantum notations given in traditional ASCII files in the static 160-character format in HITRAN.
In this notation, seemingly only eight branches are possible for the $B\,^{3}\Sigma^{-}_{u} - X\,^{3}\Sigma^{-}_{g}$ transition: 
N($N''$)P($J''$), 
P($N''$)P($J''$), 
P($N''$)Q($J''$), 
P($N''$)R($J''$), 
R($N''$)P($J''$), 
R($N''$)Q($J''$), 
R($N''$)R($J''$), and 
T($N''$)R($J''$), where $N''$ and $J''$ refer to the lower state values.    However, both HITRAN and traditional notations are both able to uniquely identify each transition  (e.g., T(5)R(6) or $^{\textrm{T}}$R$_{31}$(6) shown in Fig.~\ref{fig:levels}). Although, one has to keep in mind that some of these ``branches'' in the HITRAN notation represent more than one branch in the traditional notation, for instance, R($N''$)R($J''$) transitions can be 
$^{\textrm{R}}$R$_{33}$($J''$), 
$^{\textrm{R}}$R$_{22}$($J''$) and 
$^{\textrm{R}}$R$_{11}$($J''$). Nevertheless, since they correspond to the transitions between different spin components, they can be uniquely identified with rotational and total angular momentum quanta in the case (b) framework. Fig.~\ref{fig:levels} demonstrates all branches of the $B\,^{3}\Sigma^{-}_{u} - X\,^{3}\Sigma^{-}_{g}$ electronic transition with $N''=5$ where each individual transition has been labeled with both notations.  

In the $B''\,^{3}\Pi_u$ state, $\Omega$ can have values 0, 1, and 2, because $L$=1, also only (+) parity levels are allowed, resulting in every rotational level being populated but missing one of the $\Lambda$-doubling components. For the $B''\,^{3}\Pi_u$--$X\,^{3}\Sigma^{-}_{g}$ transition the following 15 types of branches are possible: 
N($N''$)P($J''$), 
O($N''$)P($J''$), 
P($N''$)P($J''$), 
Q($N''$)P($J''$), 
R($N''$)P($J''$), 
O($N''$)Q($J''$), 
P($N''$)Q($J''$), 
Q($N''$)Q($J''$), 
R($N''$)Q($J''$), 
S($N''$)Q($J''$), 
P($N''$)R($J''$),
Q($N''$)R($J''$), 
R($N''$)R($J''$), 
S($N''$)R($J''$), and
T($N''$)R($J''$).

The higher-lying electronic states of \ch{S2} are severely perturbed and predissociated. The predissociated bands of \ch{S2} are caused by the spin-orbit interaction with the crossing ungerade electronic states \citep{10.1063/1.5029930} and by the $B$ state crossing the dissociation limit after $v$$\geq$10 \citep{10.1063/1.470810}. Predissociation causes individual rovibronic lines to have lifetime broadening (FWHM) of up to 50~cm$^{-1}$ \citep{10.1063/1.5029930,10.1063/1.476074}. The lifetime broadening exceeds typical pressure broadening by up to three orders of magnitude and, therefore, results in a series of unresolved bands at higher frequencies. Although well described by the Lorentzian profile (as both pressure and predissociation line-widths are lifetime-driven broadening), the predissociation widths do not have the pressure and temperature dependence unlike pressure-broadened widths.

The ungerade electronic states are responsible for contributing to the predissociation. \citet{Patino1982} proposed that the primary perturber of the lower vibrational levels of the $B$ state was the $B''\,{^3}\Pi_{u}$ state. \citet{Matsumi1984,Matsumi1985} were able to measure the transitions involving $B''\,{^3}\Pi_{u}$ state directly, and Figure 2 of \citet{Matsumi1985} effectively shows the nature of perturbations.  For higher vibronic levels, \citet{10.1063/1.476074} have predicted that the $^{1}\Pi_u$ state was the primary culprit for predissociating the $v'$$\leq$16 bands and also suggested that the 1$^{5}\Pi_u$ state is responsible for predissociating the $v'$$\geq$17 bands. \citet{10.1063/1.5029930} also do not expect the $B''$ state to be the primary source of the significant predissociation for the bands $v'=11$-16. Instead, they support that theory of \citet{10.1063/1.476074} that the primary perturbers are the $^{1}\Pi_u$ and 1$\,^{5}\Pi_u$ states (for the bands $v'$$\geq$12 and $v'$$\geq$17 respectively), along with 2$\,^{3}\Sigma^{+}_u$ state for the bands $v'$$\geq$23.

There is a century-long history of experimental works that have analyzed the UV spectrum of \ch{S2} \citep{PhysRev.37.490, Olsson1936, Ikenoue1953, Meakin1962, Heaven1984, Matsumi1984, Matsumi1985, 10.1063/1.470810, 10.1039/A606591K, 10.1063/1.5029930}. Large perturbations, which cause the regular patterns within branches to break, are a consequence of the $B''$ state. Historically, these perturbations had made high-resolution analyses difficult, but \citet{10.1063/1.470810} and \citet{10.1039/A606591K} were able to provide a deperturbed rotational analysis for the $B-X$ and $B''-X$ transitions that could accurately model laser-induced fluorescence spectra. \citet{10.1063/1.476074} have investigated the predissociated bands of the $B-X$ transition as well as the perturbing levels. 
More recently, cross-sections of \ch{S2} have been measured at 370~K and 823~K using a synchrotron facility by \citet{10.1063/1.5029929}. These spectra were analyzed by \citet{10.1063/1.5029930} to develop a model for the UV spectrum of \ch{S2} that includes the $B-X$ predissociated bands. 

In each of these prior works, rotational constants, and model details have been provided, with some earlier works providing measured line positions. However, a line list that can be used directly in radiative transfer models was not available. 
Recently, an \textit{ab initio} line list  has become available \citep{sarka_2023} that can be used to simulate the spectrum of \ch{S2}. Spin-splittings and perturbation constants were not included, which has a number of limitations when used for high-resolution applications.      

\begin{figure}[ht]
    \centering
	\includegraphics[width=0.46\textwidth]{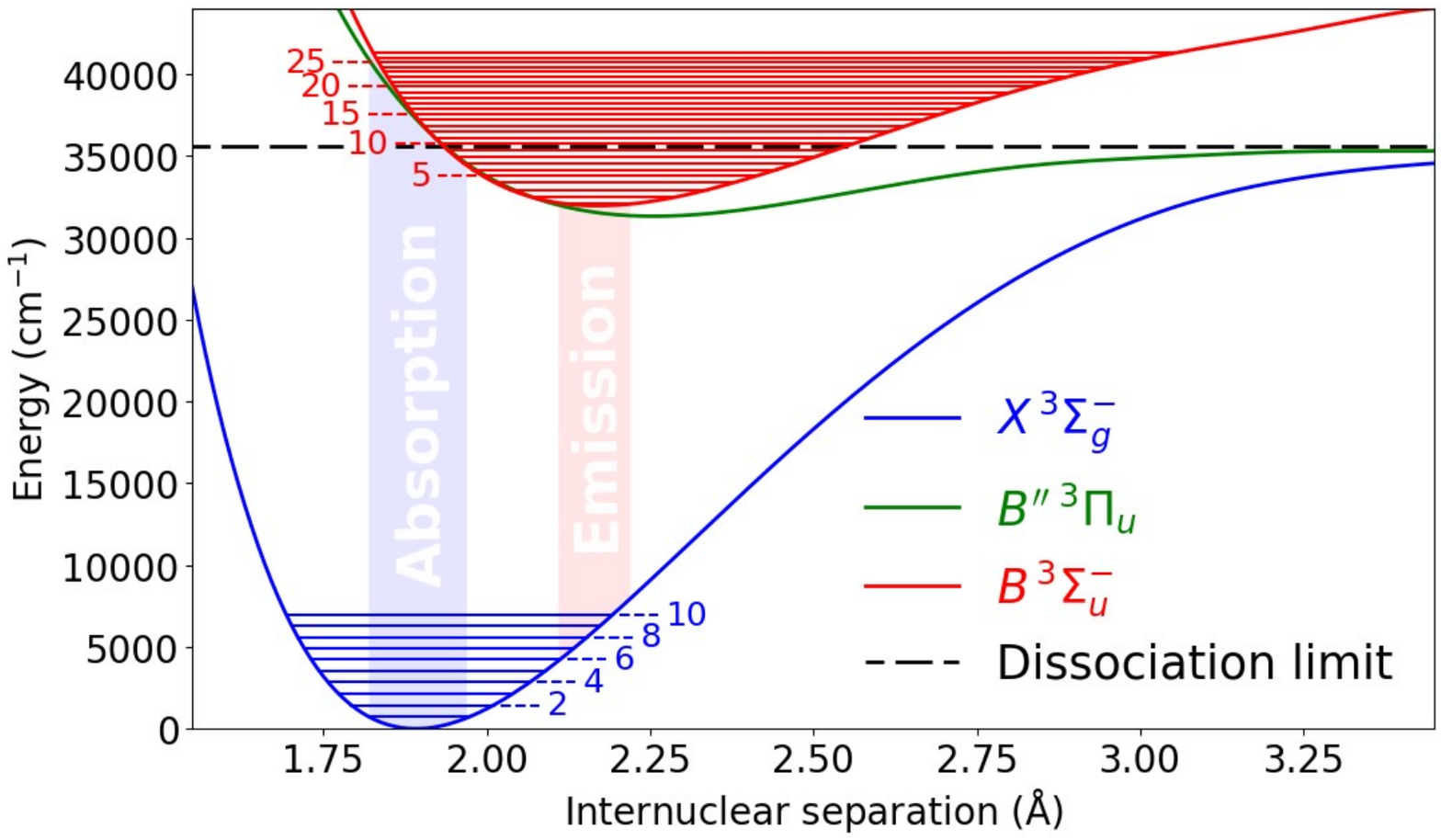}
    \caption{\ch{S2} potential energy curves for the $X$, $B$ and $B''$ states. The vibrational levels of the $X$ and $B$ states used in this work have been included, along with the dissociation limit. Shaded areas demonstrate the strongest transitions according to the Franck-Condon principle for absorption (shaded blue) and emission (shaded red) between the $X$ and $B$ states. The potential energy curves have been plotted using data from \citet{10.1016/j.jqsrt.2019.106805}. The dissociation limit of 35\,636.9 cm$^{-1}$ \citep{10.1021/acs.jpca.9b05350} is indicated, demonstrating it proximity to the $v'=10$ level of the $B$ state.}
    \label{fig:PEC}
\end{figure}

Fig.~\ref{fig:PEC} provides a schematic of the potential energy curves (PEC) of the $X$, $B$, and $B''$ states \citep{10.1016/j.jqsrt.2019.106805}. Due to the Franck-Condon principle; respective positions of the PECs; and the fact that in excited vibrational states, the overlap integrals are most efficient near the walls of the potentials, the absorption from the ground state ($v''=0$) would be most efficient to the higher vibrational states of the excited state (blue shaded region). Due to the same logic, the emission from the $v'=0$ of the $B''$ state will be most efficient to the excited vibrational levels of the ground electronic state (red-shaded region). 

%%%%%%%%%%%%%%%%%%%%%%%%%%%%%%%%%%%%%%%%%%%%%%%%%%

\section*{\small The PGOPHER \texorpdfstring{\ch{S2}}{S2} model}\vspace{-1.5mm}\hrule\vspace{3mm}\label{sec:PGO}

The \ch{S2} line list for this work has been built using the PGOPHER program \citep{10.1016/j.jqsrt.2016.04.010}, which applies spectroscopic constants to calculate transition frequencies and intensities. Our \ch{S2} model is, in large part, constructed from the analysis by \citet{10.1063/1.470810} and \citet{ 10.1039/A606591K} for rovibrational bands below the dissociation limit. In these works, laser-induced fluorescence spectra containing $B-X$ transitions of \ch{S2} were fit to a Hamiltonian that simultaneously accounted for perturbations caused by the interacting $B''$ state. In our work, some of these constants have been refit using observed emission lines of \ch{S2} \citep{Olsson1936, Ikenoue1960, Patino1982}. There are also later experimental works that reported measurements of rovibronic lines of S$_2$ \citep{Heaven1984, Matsumi1984, Matsumi1985}. However, they report only spectroscopic constants and not the actual line positions. Considering severe perturbations, it is impossible to recreate line positions from these constants without having the original program. Moreover, even with these details, these constants may not work. Anecdotally, \citet{Matsumi1984} have acknowledged that the constants provided in Table I of their paper are tentative and do not reproduce the observed line positions even at lower rotational quanta. In order to build global spectroscopic models, is imperative \citep{Gordon2016} that the experimental papers provide original measured line positions along with fitted constants. 

Table~\ref{tab-x-state} provides the term values ($T_{v}$), rotational constants ($B$), spin-spin coupling ($\lambda$), spin-rotation coupling ($\gamma$), centrifugal distortion ($D$), and the centrifugal distortion of spin-spin coupling ($\lambda_{D}$) for $v''=0$-10 of the $X\,^{3}\Sigma^{-}_{g}$ state of \ch{^{32}S_{2}} used in this work. Constants from \citet{10.1063/1.470810} were used for the $v''=0$, 1, 5 levels and PGOPHER was used to refit constants for other levels using selected line positions of the $B-X$ ($v'$,$v''$) = (2,2), (3,2), (1,3), (2,3), (3,3), (1,4), (2,4) \citep{Olsson1936}, (1,4)  \citep{Patino1982}, and (0,7), (0,8), (0,9), (0,10) \citep{Ikenoue1960} emission bands.  The $v''=6$ level was not part of the \citet{10.1063/1.470810} model, however, due to the expected importance for \ch{S2} emission in cometary spectra, we have included the $v''=6$ level by calculating $T_{6}= \omega_{e} (6.5) - \omega_{e}x_{e} (6.5)^{2} - T_{0}  = 4234.45$\,cm$^{-1}$ and $B_{v=6}=B_{e} - \alpha (6.5) +  \gamma (6.5)^{2} = 0.2851$\,cm$^{-1}$ using constants provided by \citet{Huber_Herzberg_NIST}. \citet{10.1007/BF031597601} also provide parameters to calculate $\lambda_{v=6} = 12.246$\,cm$^{-1}$ and $\gamma_{v=6} = -7.409\times10^{-3}$\,cm$^{-1}$. These have been added to our PGOPHER model to complete the $X\,^{3}\Sigma^{-}_{g}$ state for $v''=0$-10.

\begin{table*}[ht]
\caption{Deperturbed spectroscopic parameters used in the PGOPHER model for $v''=0$-10 vibrational levels of the $X\,^{3}\Sigma^{-}_{g}$ state. All values are provided in wavenumber (cm$^{-1}$) and have been refit using line positions in \citet{Olsson1936}, \citet{Ikenoue1960}, and \citet{Patino1982}. Those values unchanged from \citet{10.1063/1.470810} have been indicated.}\label{tab-x-state}
\vspace{1mm}
\centering
\begin{tabular}{ccccccc}
\hline
$v''$& $T_{v''}$  & $B$ & $\lambda$ &  $\gamma$ ($10^{-3}$) & $D$ ($10^{-7}$) & $\lambda_{D}$ ($10^{-5}$)  \\
\hline
0		&      0.0$^{a}$    &   0.2945923$^{a}$  &   11.7931$^{a}$  &  -7.157$^{a}$  &  1.96$^{a}$  &   1.05$^{a}$   \\
1		&    719.995$^{a}$  &   0.2929975$^{a}$  &   11.8659$^{a}$  &  -7.148$^{a}$  &  1.97$^{a}$  &   1.05$^{a}$   \\
2		&   1433.930        &   0.2915728        &   11.9707        &  -7.725        &  2.33        &   $-$    \\
3		&   2142.521        &   0.2899329        &   12.0575        &  -7.138        &  2.15        &   $-$    \\
4		&   2845.305        &   0.2883575        &   12.1225        &  -8.280        &  2.08        &   $-$    \\
5		&   3542.94$^{a}$   &   0.2844$^{a}$     &   12.163$^{a}$   &  -7.283$^{a}$  &  2.12$^{a}$  &   $-$    \\
6		&   4234.45$^{b}$   &   0.2851$^{b}$     &   12.246$^{b}$   &  -7.409$^{b}$  &  1.90$^{b}$  &   $-$    \\
7		&   4920.288        &   0.2827382        &   12.2883        & -34.412        & -0.359       &   $-$    \\
8		&   5600.123        &   0.2815478        &   12.3233        & -33.118        &  1.02        &   $-$    \\
9		&   6274.896        &   0.2798292        &   12.3240        & -43.742        & -0.0151      &   $-$    \\
10		&   6943.861        &   0.2779413        &   12.3744        & -50.146        & -1.38        &   $-$    \\
\hline
\multicolumn{7}{p{9cm}}{\footnotesize{$^{a}$} From \citet{10.1063/1.470810}.}  \\
\multicolumn{7}{p{9cm}}{\footnotesize{$^{b}$} Calculated from \citet{Huber_Herzberg_NIST} and \citet{10.1007/BF031597601}, see text for details.}  \\
\end{tabular}
\end{table*}

The $B\,^{3}\Sigma^{-}_{u}$ state for \ch{^{32}S_{2}} in our model includes vibrational levels from below and above the dissociation limit.  Deperturbed constants ($T_{v}$, $B$, $\lambda$, $\gamma$, $D$, $\lambda_{D}$) for the  $v'=0$-10 levels of the $B$ state have been added to our model using the values available in \citet{10.1063/1.470810} and \citet{10.1039/A606591K}. These levels mainly lie below the dissociation limit and can be resolved in experimental spectra. Table~\ref{tab-b-0-3} summarizes constants for the  $v'=0$-3 levels that were refit using PGOPHER simultaneously with those in the $X$ state using line positions from \citet{Olsson1936}, \citet{Ikenoue1960} and \citet{Patino1982}. It should be noted that \citet{10.1039/A606591K} only recommend the $v'=7$-9 constants for modeling the lower rotational levels, while the $v'=10$ constants are approximate because of difficulty modeling this rovibronic transition -- a consequence of its close proximity to the dissociation limit.

\begin{table*}
\caption{Deperturbed spectroscopic parameters used in the PGOPHER model for $v'=0$-3 vibrational levels of the $B\,^{3}\Sigma^{-}_{u}$ state. All values are provided in wavenumber (cm$^{-1}$) and have been refit to line positions in \citet{Olsson1936}, \citet{Ikenoue1960}, and \citet{Patino1982}.\label{tab-b-0-3} }
\centering
\begin{tabular}{cccccc}
\hline
$v'$& $T_{v'}$  & $B$ & $\lambda$ &  $\gamma$ ($10^{-3}$) & $D$ ($10^{-7}$)    \\
\hline
0		& 31672.413  &  0.223631     &  4.1813  &   123.425  & 16.20   \\
1		& 32102.782  &  0.223070     &  4.8517  &   -23.489  &  2.48   \\
2		& 32525.293  &  0.223487     &  2.8111  &    -8.854  &  6.89   \\
3		& 32943.385  &  0.219805     &  3.8772  &     0.173  &  2.93   \\
\hline
\end{tabular}
\end{table*}

Accurate modeling of the $B-X$ transition requires consideration of the interacting $B''\,^{3}\Pi_{u}$ state of \ch{^{32}S_{2}}. In our model, deperturbed constants ($T_{v}$, $B$, $D$, $\lambda$, $\gamma$, the spin-orbit coupling $A$, and the $\Omega=0$ lambda doubling constant $o$) are included for the $B''$ $v'=0$-11 \citep{10.1063/1.470810} and $B''$ $v'=12$-21 \citep{10.1039/A606591K} levels.

As discussed above, the $B''-B$ perturbation is essential to provide accurate positions for lines within each interacting band. The Hamiltonian model of \citet{10.1063/1.470810} accounts for perturbations of vibronic levels through interacting spin-orbit ($\alpha$) and $L$-uncoupling parameters ($\beta$), along with the  corresponding centrifugal distortion parameters ($\alpha_{D}$, $\beta_{D}$). PGOPHER allows the inclusion of the $B''-B$ interaction parameters from \citet{10.1063/1.470810} and  \citet{10.1039/A606591K}, but a difference in definition requires $\alpha$ and $\alpha_{D}$ from these work to be multiplied by $-3\sqrt{2}$. In conjunction with the PGOPHER fitting for the $X$ and $B$ states indicated above, the perturbation constants for the interacting $B''v' - Bv'=2$-0, 3-1, 4-1, 4-2, 5-2 states were also refitted using line positions from \citet{Olsson1936}, \citet{Ikenoue1960} and \citet{Patino1982} and are given in Table~\ref{tab-perturbs}. All other perturbation constants were provided by \citet{10.1063/1.470810} and \citet{10.1039/A606591K}. Overall, perturbation parameters in our model span vibrational levels $v'=0$-9 for the $B$ state, and $v'=0$-19 for the $B''$ state. As noted earlier, the difficulty in the analysis of the $B$ $v'=10$ level means that perturbation constants are unavailable.

\begin{table}
\caption{Perturbation constants (in cm$^{-1}$) that have been refit using line positions available from \citet{Olsson1936}, \citet{Ikenoue1960}, and \citet{Patino1982}.\label{tab-perturbs} }
\centering
\begin{tabular}{ccccc}
\hline
$B''v' - Bv'$ &  $\alpha$  & $\alpha_{D}$ ($10^{-3}$) & $\beta$ ($10^{-2}$)  &  $\beta_{D}$ ($10^{-5}$)   \\
\hline
$2-0$	&  -50.3926  &  4.3987   &  29.9227  &  -12.0297 \\
$3-1$	&  -61.8478  &  1.1998   &  -7.9430  &   $-$     \\
$4-1$	&   39.1613  &  1.8540   &   3.1993  &   $-$     \\
$4-2$	&   $-$      &  7.7830   &  -5.3971  &   -2.8794 \\
$5-2$	&   50.5328  &  2.8417   &   7.2030  &   $-$     \\
\hline
\end{tabular}
\end{table}

Considering those bands above the dissociation limit, \citet{10.1063/1.5029930} have built a coupled-channel model of the $B-X$ transition to account for predissociation of the $v'=11$-27 vibrational levels of the $B$ state. Term energies in \citet{10.1063/1.5029930} are provided with respect to the $F_{2}$, $v=J=0$ level of the $X\,^{3}\Sigma^{-}_{g}$  state and are separated for each $\Omega$ component (i.e., Hunds case (c) coupling was assumed). As we discussed above, this level does not actually exist for the principle isotopologue due to nuclear spin statistics, and in our PGOPHER model, this virtual state would be at an energy of 8.2~cm$^{-1}$. To implement the term values of \citet{10.1063/1.5029930} into our PGOPHER model, a calibration is required that removes the artificially applied splitting and accounts for the energy of the $F_{2}$, $v=J=0$ level of the $X\,^{3}\Sigma^{-}_{g}$ state. This essentially adds $\lambda+8.2$~cm$^{-1}$ to each $\Omega=0$ term value of \citet{10.1063/1.5029930}. Our calibrated term values are provided in Table~\ref{tab-pre-diss-bands}, along with  rotational constants and predissociation widths determined from \citet{10.1063/1.5029930}. For some bands, we have slightly adjusted the rotational constants for better agreement at higher temperatures, as indicated in Table~\ref{tab-pre-diss-bands}. \citet{10.1063/1.5029930} also recommend $\gamma=0.02$~cm$^{-1}$ for all vibrational levels of the $B$ state and this value is also included in our model for the $v'=11$-27 vibrational levels of the $B$ state. 

\begin{table}
\caption{Spectroscopic parameters used in the PGOPHER model for $v'=11$-27 vibrational levels of the $B\,^{3}\Sigma^{-}_{u}$ state. All values are provided in cm$^{-1}$ and those values that have been adjusted from \citet{10.1063/1.5029930} have been indicated.\label{tab-pre-diss-bands} }
\centering
\begin{tabular}{p{1.0cm}p{1.5cm}p{1.5cm}p{1.0cm}c}
\hline
$v'$& $T_{v'}$$^{a}$  & $B$$^{b}$ & $\lambda$$^{b}$ &  $\Gamma_{\textrm{HWHM}}$$^{c}$  \\
\hline
11		& 36\,108.4  &  0.2050        &  3.0         &  5.05  \\
12		& 36\,475.8  &  0.2028        &  0.5         &  6.35  \\
13		& 36\,841.1  &  0.2030$^{d}$  &  1.7         &  7.85  \\
14		& 37\,200.1  &  0.2005        &  3.3        &  5.75  \\
15		& 37\,551.4  &  0.1970$^{d}$  &  3.4         &  4.15  \\
16		& 37\,895.8  &  0.1945$^{d}$  &  2.5         &  3.30  \\
17		& 38\,232.3  &  0.1920$^{d}$  &  1.7         &  3.45  \\
18		& 38\,560.1  &  0.1909        &  1.3         &  8.80  \\
19		& 38\,889.0  &  0.1921        &  2.8         & 25.05  \\
20		& 39\,233.8  &  0.1908        &  5.0         & 24.85  \\
21		& 39\,539.6  &  0.1796        &  5.3         &  3.20  \\
22		& 39\,850.1  &  0.1876        &  3.6         & 17.35  \\
23		& 40\,143.7  &  0.1765        &  4.9         &  5.25  \\
24		& 40\,445.1  &  0.1793        &  4.6         &  9.45  \\
25		& 40\,721.7  &  0.1782        &  3.2         & 14.45  \\
26		& 40\,998.2  &  0.1696        &  4.3         &  3.15  \\
27		& 41\,277.1  &  0.1718        &  7.9         &  6.20  \\
\hline
\multicolumn{5}{p{8cm}}{\footnotesize{$^{a}$} Term values have been calibrated from the separated $\Omega$ components in \citet{10.1063/1.5029930}. See text for details.} \\
\multicolumn{5}{p{8cm}}{\footnotesize{$^{b}$} Calculated values from \citet{10.1063/1.5029930}, unless otherwise stated.} \\
\multicolumn{5}{p{8cm}}{\footnotesize{$^{c}$} Based on the average calculated values in \citet{10.1063/1.5029930}.} \\
\multicolumn{5}{p{8cm}}{\footnotesize{$^{d}$} Constant has been refit for this work.}
\end{tabular}
\end{table}

%%%%%%%%%%%%%%%%%%%%%%%%%%%%%%%%%%%%%%%%%%%%%%%%%%
\subsection*{\small Refitting rotational constants for emission bands}\vspace{-1.35mm}\hrule\vspace{3mm} \label{sec:constantfitting}

\citet{10.1063/1.470810} provided a deperturbed analysis of the $B - X$ transition of \ch{S2} with $v'=0$-6, $v''=0$-7, and the inclusion of perturbations from $v'=2$-12 of the $B''$ state . Their work used a combination of line positions observed from laser-induced fluorescence spectra (primarily from lower $J$ levels) and previously recorded high-temperature static cell measurements (including levels with $J$ up to 100). In total, 3320 observed lines went into the fitting of their initial \ch{S2} model. 

Applying the rotational constants of \citet{10.1063/1.470810} into an initial model for this work, comparisons were made to the measured positions for the $B-X$ ($v'$,$v''$) = (2,2), (3,2), (1,3), (2,3), (3,3), (1,4), (2,4) bands from \citet{Olsson1936}, the (1,4) band from \citet{Patino1982}, and the (0,7) band from \citep{Ikenoue1960}. Mismatched assignments due to band crossings were removed, however there remained differences of up to $~2$~cm$^{-1}$ to the remaining 864 observed line positions. Overlaying the output of our initial model to the fluorescence spectrum of the (3,3) band recorded by \citet{Heaven1984} implied that the lower $J$ levels were in good agreement. Therefore, to better match the observations of \citet{Olsson1936, Patino1982, Ikenoue1960} some modeled lines were included in the refitting of the rotational constants from \citet{10.1063/1.470810} to restrain the fitted parameters. The first five lines of the $^{\textrm{R}}$R$_{11}$($J''$),  $^{\textrm{R}}$R$_{22}$($J''$),  $^{\textrm{R}}$R$_{33}$($J''$),  $^{\textrm{P}}$P$_{11}$($J''$), $^{\textrm{P}}$P$_{22}$($J''$), and $^{\textrm{P}}$P$_{33}$($J''$) branches of the (2,2), (3,2), (1,3), (2,3), (3,3), (1,4), (2,4), and (0,7) bands were included in the fit using the line positions of our initial model \citep[i.e., predicted using constants from][]{10.1063/1.470810}. This effectively limited the impact of the refitting for the lower $J$ levels of these bands, while giving enough flexibility of the fit to account for the measurements of \citet{Olsson1936, Patino1982, Ikenoue1960}. It was also necessary to fit the perturbation constants as they have a significant influence on the resultant spectrum. The refitting of the constants for these bands yielded a standard deviation of 0.133 cm$^{-1}$. 

\citet{Ikenoue1960} also included measured positions for the (0,8), (0,9), and (0,10) bands. This allowed the rotational constants of the $v''=8$, 9, and 10 levels to be predicted, which were not available in \citet{10.1063/1.470810}. These bands were fit using 266 line position from \citet{Ikenoue1960} to give a standard deviation of 0.116~cm$^{-1}$.

In total, 1378 lines were used for a standard deviation of 0.130~cm{$^{-1}$}. The constants for levels of the $B-X$ transition that were fit for this work are provided in Tables~\ref{tab-x-state}, \ref{tab-b-0-3}, and \ref{tab-perturbs}. 

%%%%%%%%%%%%%%%%%%%%%%%%%%%%%%%%%%%%%%%%%%%%%%%%%%

\subsection*{\small Determining band strengths for predissociated region}\vspace{-1.5mm}\hrule\vspace{3mm} \label{sec:FVVvalues}

Theoretical oscillator strengths ($f_{v'v''}$) are reported in the literature for many bands of \ch{^{32}S_{2}} \citep{10.1016/0009-2614(96)00363-6, 10.1016/0022-4073(71)90160-9}. However, these works do not cover all of the $B-X$ predissociated bands measured by \citet{10.1063/1.5029929} and often do not include hot bands.  The Einstein-A coefficients reported by \citet{10.1016/j.jqsrt.2019.106805} have been converted to oscillator strengths using the formulae in \citet{Bernath_book}. In addition, some works \citep{10.1063/1.438372,  10.1007/s10509-019-3656-3, 10.1016/j.jqsrt.2019.106805} report the Frank-Condon (FC) factors ($q_{v'v''}$), but these have not been implemented in this work. 

\begin{figure}[ht]
\centering
	\includegraphics[width=8cm]{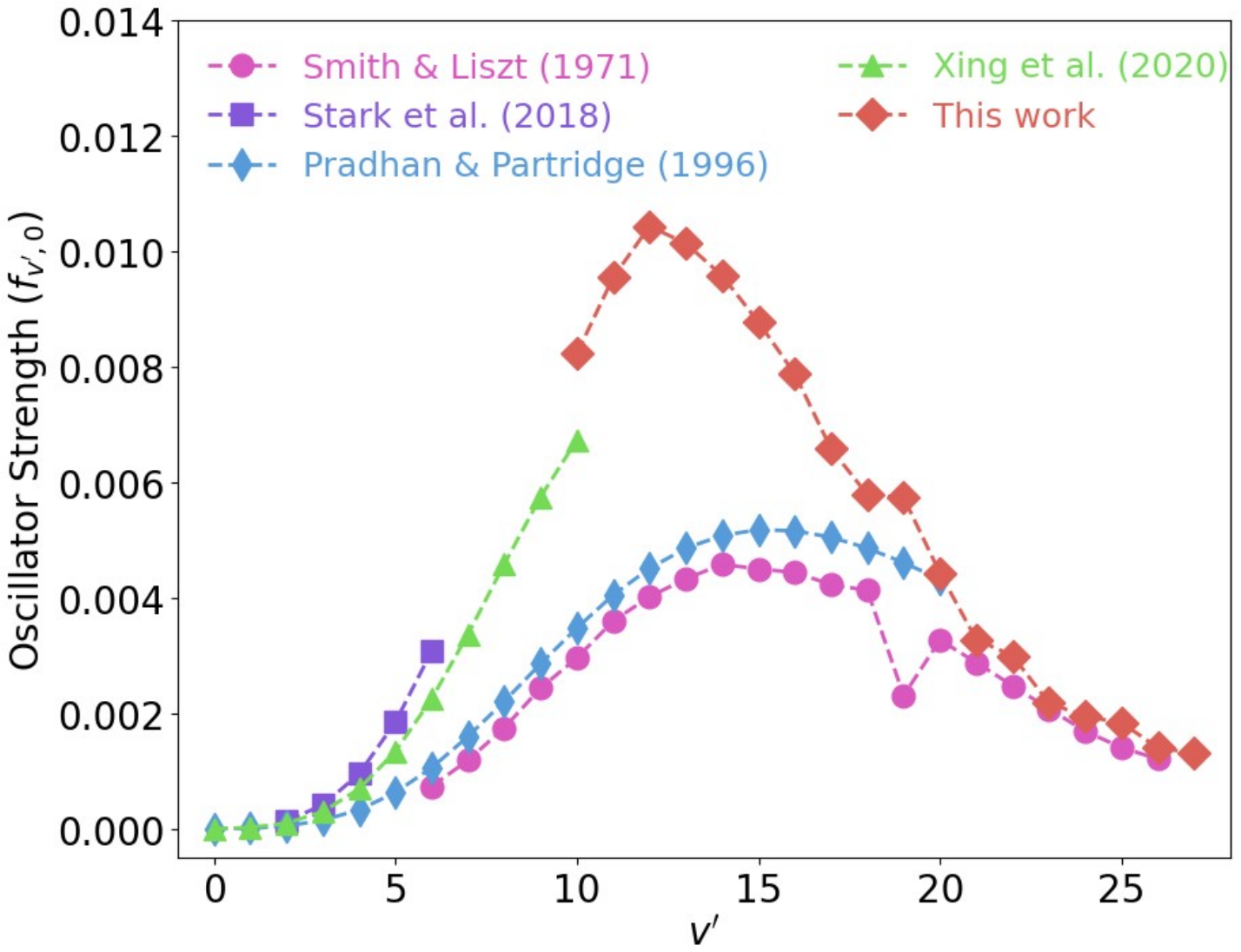}
    \caption{Oscillator strengths ($f_{v',0}$) factors for the bands of the $B-X$ transition of \ch{^{32}S_{2}}, plotted against $v'$ with $v''=0$ \citep{10.1016/0022-4073(71)90160-9, 10.1063/1.5029929, 10.1016/0009-2614(96)00363-6, 10.1016/j.jqsrt.2019.106805}. Also included are oscillator strengths from this work that were obtained by fitting to experimental absorption cross sections at 370 and 823~K from  \citet{10.1063/1.5029929}.}
    \label{fig:FC}
\end{figure}

Fig.~\ref{fig:FC} provides a comparison of the oscillator strengths considered in this work for $B-X$  bands with $v''=0$. Generally, there is a good agreement between \citet{10.1063/1.5029929} and \citet{10.1016/j.jqsrt.2019.106805} for fundamental bands with $v'<10$. However, since the  the oscillator strengths of \citet{10.1016/0009-2614(96)00363-6} and \citet{10.1016/0022-4073(71)90160-9} do not align well with those of \citet{10.1063/1.5029929}, it was necessary to fit the oscillator strengths using PGOPHER \citep{10.1016/j.jqsrt.2016.04.010} for the $v'\geq10$ fundamental bands using the calibrated cross-sections at 370 and 823~K \citep{10.1063/1.5029929}.

PGOPHER requires the square root of the band strength (i.e, $\sqrt{S}$\footnote{where $S \propto q_{v'v''}|R^{2}|$ and $R$ is the transition dipole moment}) in order to scale the strength of the corresponding transition and individual line intensities. These can be converted to oscillator strengths using the formulae in \citet{Bernath_book}. The resultant oscillator strengths determined from the fit are shown in  Fig.~\ref{fig:FC} and provided in Table~\ref{tab-fvv}. The fitted values appear consistent with those from \citet{10.1063/1.5029929} and \citet{10.1016/j.jqsrt.2019.106805} for the $v'<10$ bands and also are in qualitative agreement when compared to the oscillator strengths for the predissociated region presented in Figure 6 of \citet{10.1063/1.5029929}. Our fit also indicates that the band with the strongest overlap (i.e., maximum oscillator strength) appears at $v'=12$, which is a slightly lower vibrational level than that of \citet{10.1016/0009-2614(96)00363-6} and \citet{10.1016/0022-4073(71)90160-9}, but consistent with the vibrational level reported by \citet{10.1063/1.5029929}.

Hot bands with $v''=1$ are also prominent in the \citet{10.1063/1.5029929} spectra, particularly at 823~K, therefore it was also necessary to fit the oscillator strengths for $v'\geq13$ for these bands. The fitted oscillator strengths for the hot bands are also given in Table~\ref{tab-fvv}.

For $B-X$ bands with $v''=0$ below the dissociation limit ($v'$=0-9), the oscillator strengths calculated from \citet{10.1016/j.jqsrt.2019.106805} have been used, given the consistency with \citet{10.1063/1.5029929} for the $v''=0$ bands. Oscillator strengths calculated from  \citet{10.1016/j.jqsrt.2019.106805} were also used for the $v''=1$ and $v''=2$  hot bands up to $v'$=10. The oscillator strengths for the $v'=11$-12 bands have been estimated due to the lack of consistency between literature values. Oscillator strengths for the $v''=2$ hot bands with $v'$=11-20 are provided by \citet{10.1016/0009-2614(96)00363-6}, with an estimated strength for the $v'$=21-27 bands. 

A summary of the oscillator strengths used in this work for the $B-X$ bands with $v''=0$-2 is provided in Table~\ref{tab-fvv}. For $B-X$ bands with $v'=0$-3 and $v''=3$-10, oscillator strengths calculated from \citet{10.1016/j.jqsrt.2019.106805} have been used.

\begin{table}
\caption{Oscillator strengths ($f_{v',v''}$) used in this work for \ch{^{32}S_{2}}. These include those obtained from fits to experimental cross-sections at 370 and 823~K  \citep{10.1063/1.5029929}, and when indicated, those taken from the literature. \label{tab-fvv} }
\centering
\begin{tabular}{llll}
\hline
$v'$ & $f_{v',0}$ ($10^{2}$)  & $f_{v',1}$ ($10^{2}$) & $f_{v',2}$ ($10^{2}$)  \\
\hline
0&       0.00028$^{a}$&       0.00322$^{a}$&       0.01818$^{a}$ \\  
1&       0.00231$^{a}$&       0.02202$^{a}$&       0.09851$^{a}$ \\  
2&       0.00993$^{a}$&       0.07709$^{a}$&       0.26882$^{a}$ \\  
3&       0.02971$^{a}$&       0.18455$^{a}$&       0.48764$^{a}$ \\  
4&       0.06932$^{a}$&       0.33894$^{a}$&       0.65007$^{a}$ \\  
5&       0.13428$^{a}$&       0.50620$^{a}$&       0.66086$^{a}$ \\  
6&       0.22537$^{a}$&       0.63762$^{a}$&       0.51167$^{a}$ \\  
7&       0.33722$^{a}$&       0.69030$^{a}$&       0.28159$^{a}$ \\  
8&       0.45872$^{a}$&       0.64932$^{a}$&       0.08544$^{a}$ \\  
9&       0.57501$^{a}$&       0.52711$^{a}$&       0.00129$^{a}$ \\  
10&       0.82326      &       0.36166$^{a}$&       0.04160$^{a}$ \\  
11&       0.95497      &       0.18455$^{b}$&       0.00300$^{c}$ \\  
12&       1.04341      &       0.07709$^{b}$&       0.01200$^{c}$ \\  
13&       1.01371      &       0.27715      &       0.06200$^{c}$ \\  
14&       0.95778      &       0.35418      &       0.12800$^{c}$ \\  
15&       0.87753      &       0.40679      &       0.19200$^{c}$ \\  
16&       0.78794      &       0.50050      &       0.23800$^{c}$ \\  
17&       0.65796      &       0.58404      &       0.26200$^{c}$ \\  
18&       0.57876      &       0.81386      &       0.26200$^{c}$ \\  
19&       0.57341      &       0.94788      &       0.24200$^{c}$ \\  
20&       0.44291      &       0.75558      &       0.21000$^{c}$ \\  
21&       0.32794      &       0.52221      &       0.19053$^{b}$ \\  
22&       0.29978      &       0.69719      &       0.17074$^{b}$ \\  
23&       0.21819      &       0.61324      &       0.15054$^{b}$ \\  
24&       0.19473      &       0.55452      &       0.13004$^{b}$ \\  
25&       0.18400      &       0.57918      &       0.10913$^{b}$ \\  
26&       0.14052      &       0.48191      &       0.08792$^{b}$ \\  
27&       0.13119      &       0.48519      &       0.06641$^{b}$ \\  
\hline
\hline
\multicolumn{4}{p{5cm}}{\footnotesize{$^{a}$} Calculated from \citet{10.1016/j.jqsrt.2019.106805}.} \\
\multicolumn{4}{p{5cm}}{\footnotesize{$^{b}$} Estimated values.}\\
\multicolumn{4}{p{5cm}}{\footnotesize{$^{c}$}  \citet{10.1016/0009-2614(96)00363-6}.}
\end{tabular}
\end{table}

%%%%%%%%%%%%%%%%%%%%%%%%%%%%%%%%%%%%%%%%%%%%%%%%%%

\subsection*{\texorpdfstring{\small \ch{S2}}{S2} line list for HITRAN}\vspace{-1.5mm}\hrule\vspace{3mm}\label{sec:finalLL}
The line list generated for \ch{^{32}S_2} from PGOPHER has been converted into the standard format used by the HITRAN database \citep{10.1016/j.jqsrt.2021.107949}. This format is used as input to numerous radiative transfer codes for terrestrial and exoplanetary applications, and is expected to be included into the HITRAN database (for this work, 58 is used as a provisional molecule ID number). Line positions (in cm$^{-1}$), intensities (cm/molecule at 296~K), Einstein-$A$ values (s$^{-1}$), lower-state energies (cm$^{-1}$), and transition assignments have been converted directly from PGOPHER.  It is necessary to include pressure broadening parameters for each line so that spectra can be calculated from the line list. For this work, air- and self-broadening Voigt parameters have been estimated as 0.05 cm$^{-1}$/atm for both. In addition, the temperature dependence of the broadening has been estimated as 0.71. These values have been approximated based on comparisons to similar parameters in HITRAN for O$_{2}$. The line intensities in HITRAN formalism are scaled due to the ``natural'' terrestrial abundance of atomic species taken from \citet{10.1063/1.555720}. Therefore, the intensities in PGOPHER for 100\% \ch{^{32}S_2} are scaled by  0.9028 in the HITRAN-formatted line list.

\begin{figure}[ht]
    \centering
	\includegraphics[width=8cm]{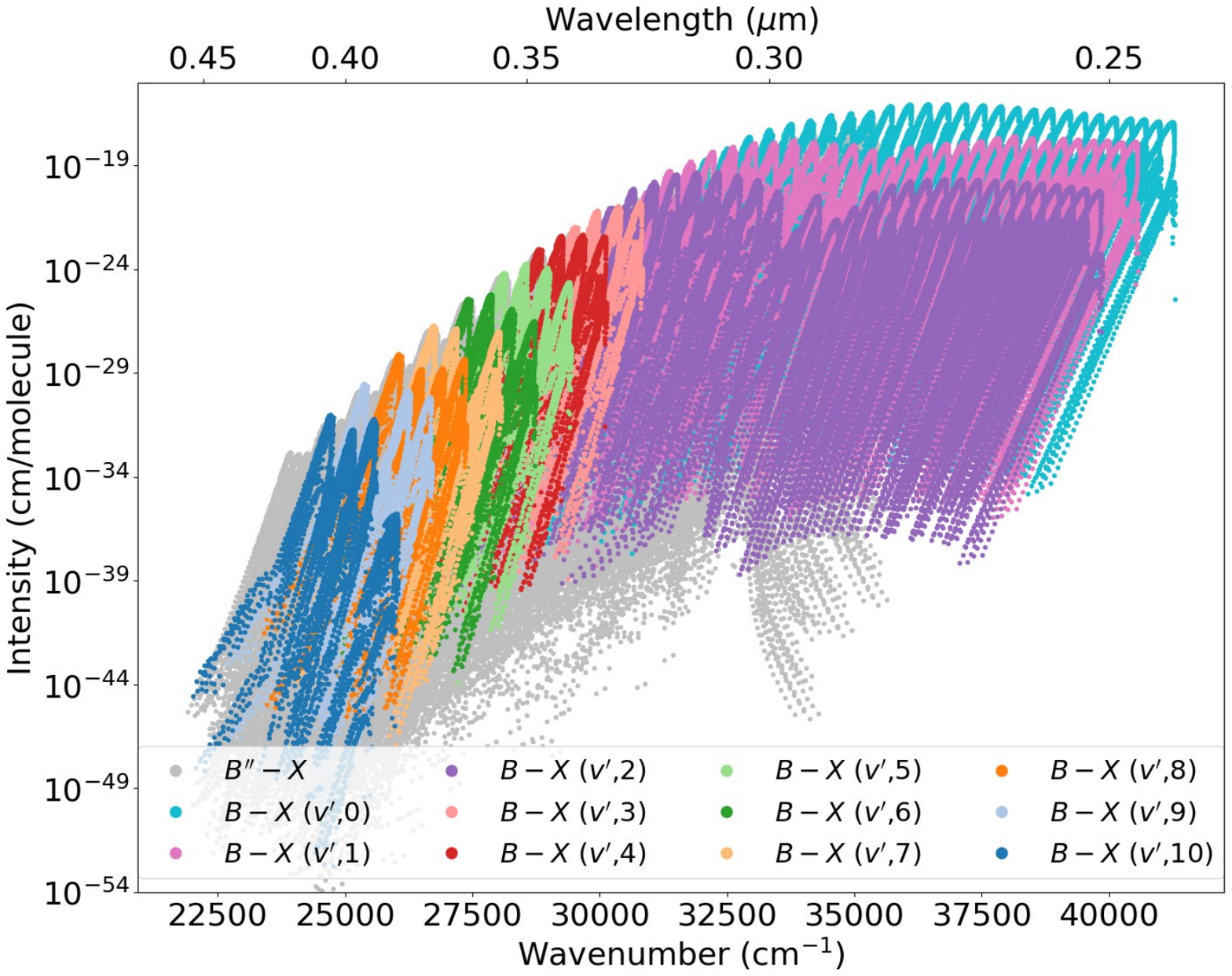}
    \caption{An overview of the \ch{S2} line list calculated for this work. Vibronic bands of the $B-X$ transition (with $v'$=0-27, $v''$=0-10) are indicated. The line positions (cm$^{-1}$) have been plotted against intensity (cm/molecule at 296~K) scaled by their natural abundance for \ch{^{32}S_2}.}
    \label{fig:linelist}
\end{figure}
An overview of the \ch{S2} line list has been provided in Fig.~\ref{fig:linelist}. The line list includes $B-X$ bands up to $v'=27$ and spans the 21\,700$-$41\,300~cm$^{-1}$ ($\sim$242$-$461~nm) spectral range.  The vertical axis shows intensity in HITRAN units and formalism (at 296~K), which assumes local thermal equilibrium. That is why the bands with high $v''$ values appear so weak, as these levels will have a very small population at 296~K. However, as shown in Figure \ref{fig:PEC}, one should expect strong emissions down to these levels from photochemically excited lower levels of the excited electronic state. 

In addition to the line list, a partition sum, $Q$, is also required to recalculate line intensities at different temperatures. For this work, the total internal partition sums ($Q_{\textrm{sum}}$) for \ch{^{32}S2} has been exported from the PGOPHER model, which employs direct summation of the energy levels. For consistency with other molecules in HITRAN, the $Q_{\textrm{sum}}$ values have been placed on the same temperature grid used in TIPS-2021 \citep{10.1016/j.jqsrt.2021.107713}. In addition, the lower state energies have been adjusted to the energy of the lowest occupied level (i.e., +15.120664\,cm$^{-1}$ has been added to all energies) to be consistent with HITRAN format, and the partition sum exported from PGOPHER.  The \ch{S2} HITRAN metadata and partition sum can be seamlessly implemented into the HITRAN Application Programming Interface, HAPI \citep{10.1016/j.jqsrt.2016.03.005}, to enable the calculation of cross sections for this work. The HITRAN-formatted \ch{S2} line list is provided as a supplementary file, along with the partition sum to allow calculation with HAPI. 

As noted earlier, the \ch{S2} diffuse $B-X$ bands in the UV require the inclusion of predissociated line widths in order to generate reliable cross-sections. \citet{10.1063/1.5029930} provides separate calculated predissociation widths for each $\Omega$ for the $v'$$\geq$10 $B-X$ bands. These have been averaged for each vibrational level and are provided as half width at half maximum (HWHM) values in Table~\ref{tab-pre-diss-bands}. A Python code has been generated to be used alongside HAPI in order to calculate absorption cross-sections with the inclusion of predissociated line widths. This code is provided as a Supplementary file and is expected to be incorporated into future versions of HAPI. It is anticipated that it will also be of benefit to the calculation of predissociation for other molecules in HITRAN, in particular for the Schumann-Runge bands of \ch{O2}. 

%%%%%%%%%%%%%%%%%%%%%%%%%%%%%%%%%%%%%%%%%%%%%%%%%%

\section*{\small Results}\label{sec:results}\vspace{-1.5mm}\hrule\vspace{3mm}

\begin{figure}[ht]
    \centering
	\includegraphics[width=8cm]{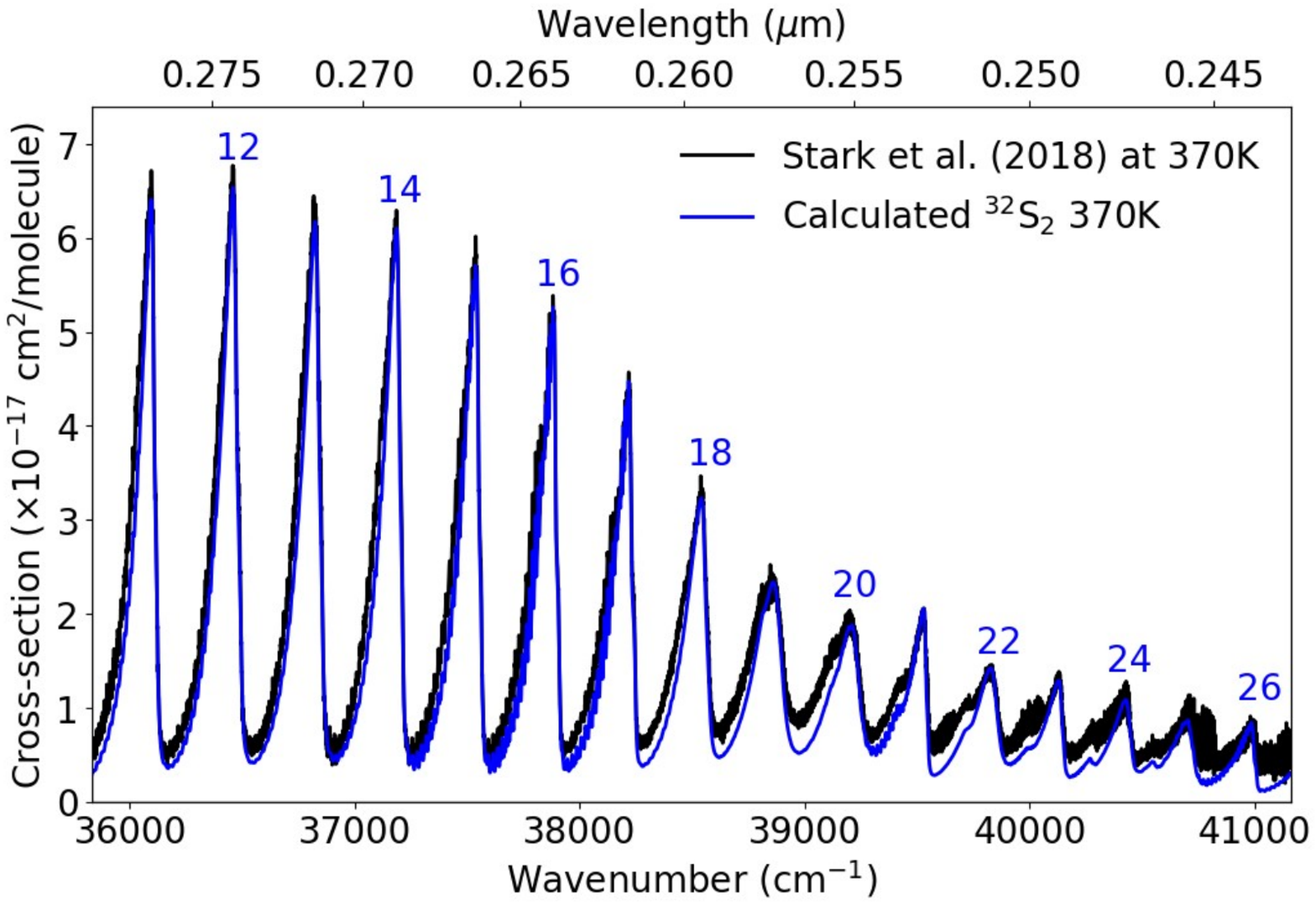} \\
	\includegraphics[width=8cm]{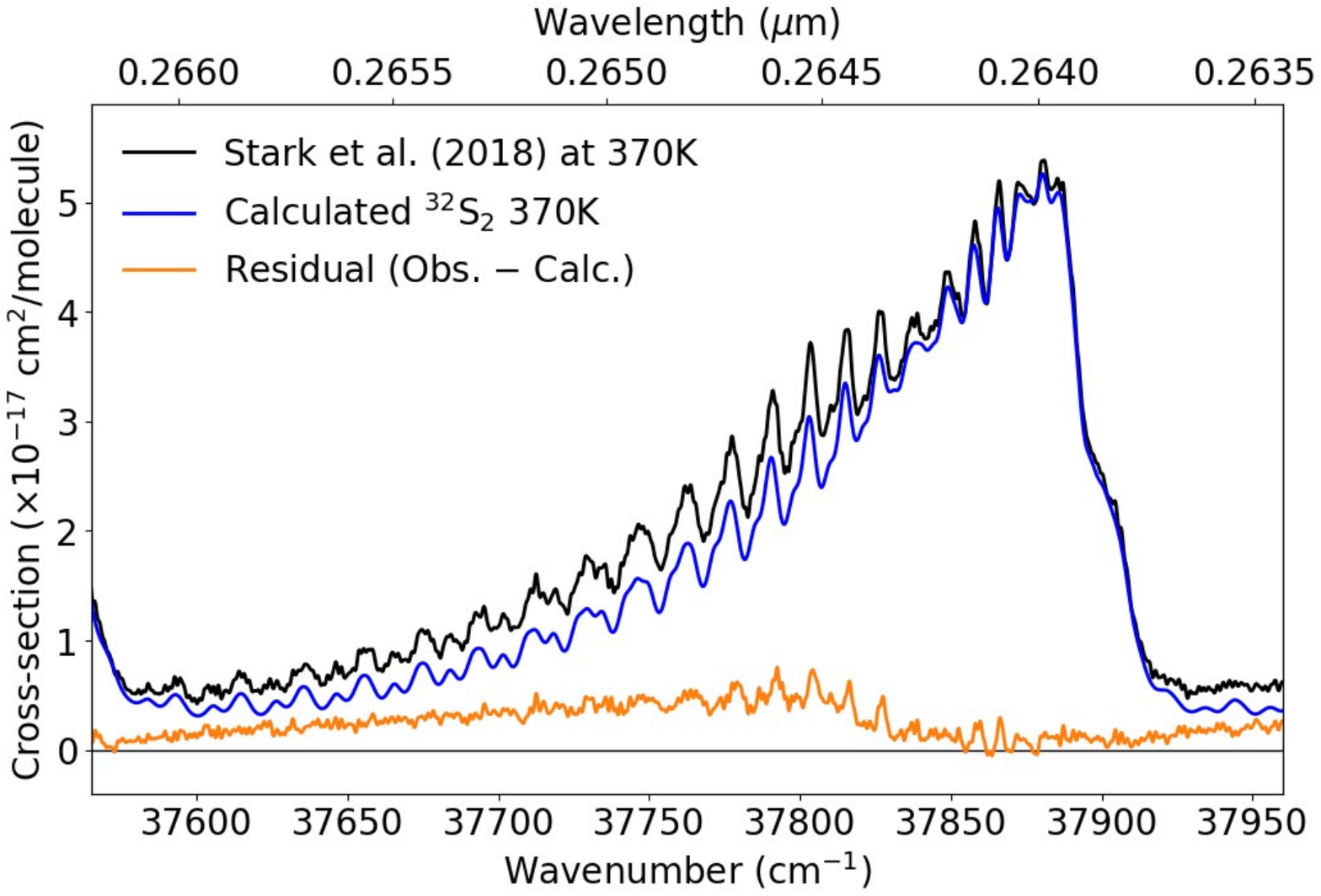} \\
    \caption{Experimental \ch{S2} absorption cross section at 370~K from \citet{10.1063/1.5029929}  compared to a calculated spectrum (blue) that uses the line list from this study. The upper panel shows an overview of the predissociated region between 35\,400-41\,500\,cm$^{-1}$ with even $v'$ indicated for the $B-X$ ($v'$,0) bands. 
    The lower panel shows a zoomed-in region around the $B-X$ (16,0) band as it has the smallest predissociated width, with a residual (Obs.-Calc.) shown in orange. 
    Predissociated bands with $v'$$\geq$10 have been calculated using the Lorentz HWHM values given in Table~\ref{tab-pre-diss-bands}. }
    \label{fig:370D}
\end{figure}

The line list generated in this work has been validated by comparing calculated spectra to the measurements of \citet{10.1063/1.5029929} at 370 and 823~K, as was done by \citet{10.1063/1.5029930}. All spectra in this work have been calculated using HAPI with the inclusion of the presdissociation widths.
The upper panel of Fig.~\ref{fig:370D} shows the \ch{S2} line list calculated at 370~K and plotted against the calibrated cross-section from \citet{10.1063/1.5029929}. 
The lower panel of Fig.~\ref{fig:370D} shows a zoomed-in region of the upper panel around the $v'=16$ band of the $B-X$ transition with the  residual (Obs.-Calc.) also shown. The $B-X$ bands above the dissociation limit exhibit a similar structure, but the size of the predissociation widths can obscure much of the detail. Spin-spin splitting leads to a separation of each $\Omega$ component and, depending on the magnitude of the separation and predissociation widths, the $\Omega=0$ branches can be seen as a shoulder to the bandhead as shown for the $v'=16$ band near 37\,900\,cm$^{-1}$. In addition, the predissociation widths for this level are small enough to resolve partial rotational structure, which agrees very well with the measurements of \citet{10.1063/1.5029929}. Some of the calculated intensity in this band is due to the underlying hot bands. 

\begin{figure}[ht]
    \centering
	\includegraphics[width=8cm]{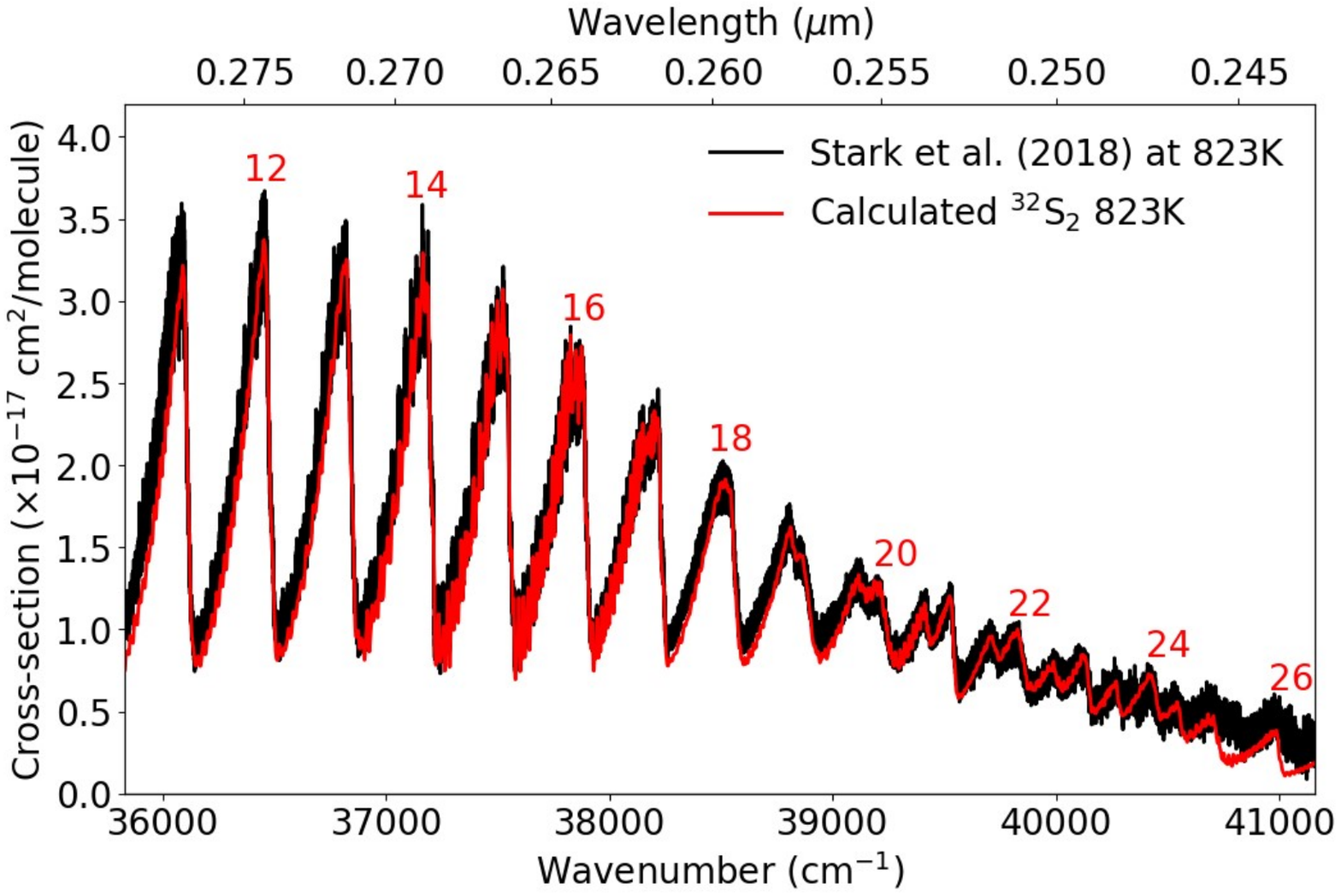} \\
	\includegraphics[width=8cm]{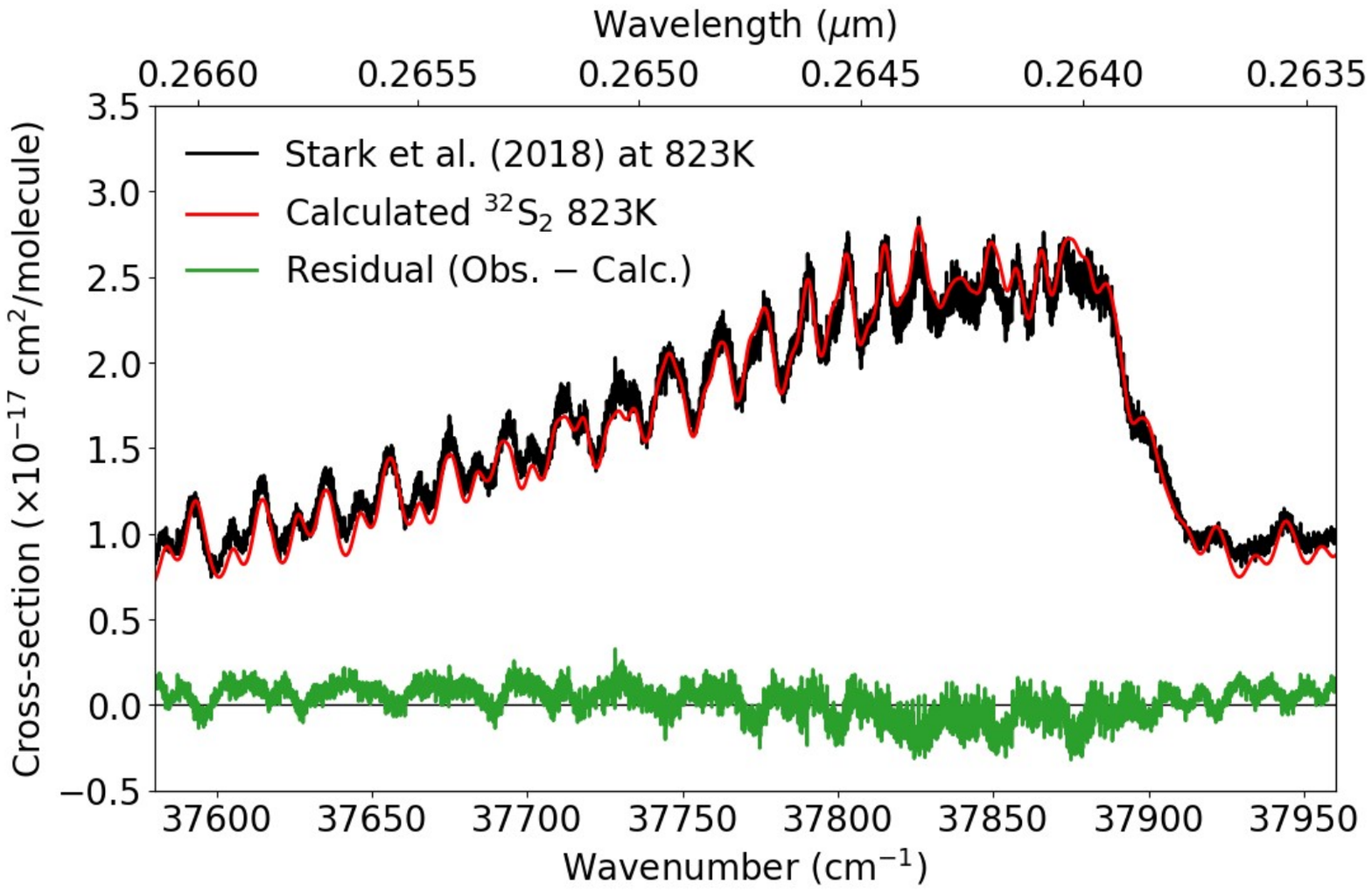} \\
    \caption{Experimental \ch{S2} absorption cross section at 823~K from \citet{10.1063/1.5029929} compared to a calculated spectrum (red) using line list from this work. The upper panel shows an overview of predissociated region between 35\,400-41\,500\,cm$^{-1}$, with even $v'$ indicated for the $B-X$ ($v'$,0) bands. Hot bands have not been indicated, but are prominent in between the ($v'$,0) sequence. The lower panel shows a zoomed-in region around the (16,0) band. This band has the smallest predissociated width, which partially reveals the rotational structure. The residual (Obs.-Calc.) is shown in green.} 
    \label{fig:823}
\end{figure}
For this work, hot bands to the $v''=1$ and $v''=2$ of the $X$ state have been included. These bands become more dominant at higher temperatures, and 
Figure~\ref{fig:823} shows comparisons to the 823~K cross section of \citet{10.1063/1.5029929}. The oscillator strengths for our model have been determined by comparing the intensities at both 370 and 823~K. At this higher temperature, good agreement is also seen with the spectrum calculated using the line list from this work. 

%%%%%%%%%%%%%%%%%%%%%%%%%%%%%%%%%%%%%%%%%%%%%%%%%%

\section*{\small Discussion}\vspace{-1.5mm}\hrule\vspace{3mm}
The primary challenge in this work was to combine the constants and parameters of previous works into a consistent model that can be used for calculating the spectrum of S$_{2}$ in the UV. Since the spectral range of the $B-X$ line list spans the dissociation limit of S$_{2}$, there is a different classification of the accuracy of individual spectroscopic parameters above and below this limit. Below the dissociation limit for bands with $v'\leq9$, the perturbation model substantially improves the accuracy of the line positions, as demonstrated by the refitting of the rotational constants. However, for bands above the dissociation limit with $v'\geq11$, the position accuracy is difficult to estimate as the lines are broadened due to predissociation, and perturbations for these bands are not included. It should be expected that these bands are consistent with those of \citet{10.1063/1.5029930}, and have a conservatively estimated position accuracy of $\sim$10~cm$^{-1}$. 

\begin{figure}[ht]
    \centering
	\includegraphics[width=8cm]{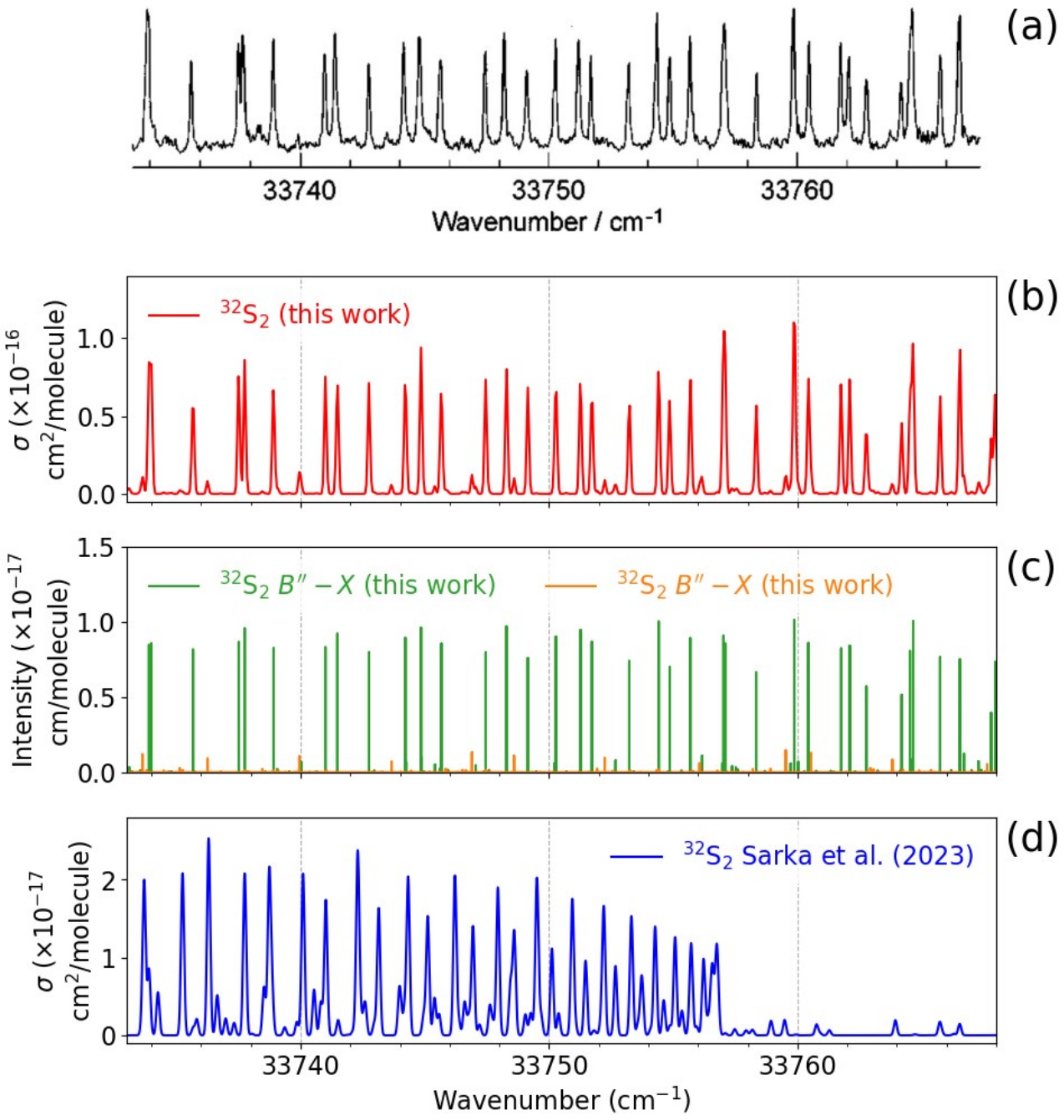}
    \caption{Comparison in the region of the $B-X$ (5,0) band of \ch{S2}. (a) Laser induced fluorescence spectrum of \ch{S2} with rotational temperature of $\sim$300~K (reprinted from Fig.~6 of \citet{10.1063/1.470810}, with the permission of AIP Publishing). (b) Experimental absorption cross section at 300~K calculated using the line list from this work and convolved with a  0.1~cm$^{-1}$ Gaussian lineshape. (c) Line positions and intensities from the HITRAN-formatted line list of this work showing the $B-X$ and weaker $B''-X$ lines. (d) Experimental absorption cross section at 300~K from the Supplementary Material of \citet{sarka_2023} convolved with a  0.1~cm$^{-1}$ Gaussian lineshape.   }
    \label{fig:GWcomp}
\end{figure}

For bands involving levels below predissociation, a standard deviation of all fitted lines is 0.130~cm$^{-1}$, which is much higher than above the predissociaion. Figure~\ref{fig:GWcomp} includes a comparison of the line list and calculated cross section from this work (at 300~K) to the measurements of \citet{10.1063/1.470810} in the region of the $B-X$ (5,0) transition. While the intensity of the experiment is not equivalent to the absorption cross section, the agreement of the line positions is excellent and demonstrates the accuracy of these bands. Strong perturbations of the $B-X$ (5,0) band with the $B''-X$ (10,0) have a significant impact on the line positions. Moreover, the $\Omega$ splitting also impacts the line positions and significantly alters the location of each band head. Including these effects is essential for reproducing experimental observations as demonstrated in Figure~\ref{fig:GWcomp} for the comparison to the absorption cross section from \textit{ab initio} work of \citet{sarka_2023}, which does not account for spin-splitting or perturbations. Furthermore, the predissociation widths included in \citet{sarka_2023} does not agree with the predissociation broadening observed in the experiments of \citet{10.1063/1.5029929}. 

It should be highlighted that only a limited number of line positions could be used when refitting the constants. Rotational constants have primarily been provided by previous works, which allows for a model to be constructed that works reasonably well for unperturbed levels. However, for a molecule like \ch{S2} where large interactions substantially perturb the levels and observed line positions, it is necessary to include as many observed transitions as possible to refine the crossing points for the $B''-B$ interaction. It was also noted that our fit for $B-X$ ($v',v''$) = (2-2), (2-3), (2-4) bands to line positions from \citet{Olsson1936} showed the largest deviations, with up to $\sim$1~cm$^{-1}$ residuals, while other bands observed by \citet{Olsson1936} performed well. Therefore we attribute this deviation to the rotational levels of the $B$~$v'$=2 state, which would need further experimental measurements to validate the rotational constants and perturbation parameters.  One of the major uncertainties of our model is the accuracy of the $B$~$v'$=10 level which only has a partially resolved rotational structure in the experimental spectra. We used constants provided in \citet{10.1039/A606591K}, but it should be stressed that the values are only estimates and do not contain any of the perturbation considerations of lower bands. Therefore, it should be expected that the line position accuracy of the model for the $B$~$v'$=10 level is closer to that of the predissociated bands.

The intensities in this work have been calculated from oscillator strengths, with some determined from fits to the experimental measurements of  \citet{10.1063/1.5029929}. Those oscillator strengths obtained from fitting have assumed that the measured spectra are primarily a consequence of the $B-X$ transition of \ch{^{32}S_{2}}. However, it would be expected that approximately 8\% of the absorption will be caused by the second-most abundant isotopologue, \ch{^{32}S^{34}S}. There is only limited spectroscopic information in the literature for the \ch{^{32}S^{34}S} isotopologue \citep[e.g.,][]{10.1039/A606591K}, and it was not sufficient to build a comprehensive line list of sufficient accuracy for this work. \citet{10.1063/1.5029930} included \ch{^{32}S^{34}S} as part of their model with a band structure consistent with the residuals seen in Figs.~\ref{fig:370D} and \ref{fig:823}. We can, therefore attribute the majority of these residuals to the \ch{^{32}S^{34}S} isotopologue and note that our \ch{^{32}S_{2}} intensities will have an increased uncertainty ($\sim$10\%) due to the absence of \ch{^{32}S^{34}S} in our model.   In addition, the oscillator strengths used in this work for the hot bands for the predissociated region are limited due to the lack of coverage in the literature, and a higher uncertainty can be expected. \citet{10.1063/1.5029929} and \citet{10.1063/1.5029930} noted an apparent continuum in the experimental spectra with an approximate intensity of $\sim$3$\times$10$^{-18}$ cm$^{2}$, which has not been included in our model and contributes to the residual (noticeable at higher wavenumbers). Moreover, there are resolved transitions in the \citet{10.1063/1.5029929} spectra near 40\,000~cm$^{-1}$ and 40\,800~cm$^{-1}$ at 370~K that are due to the $f\,^{1}\Delta_{u}-a\,^{1}\Delta_{g}$ transition, which are not included in our line list. 

Further experimental measurements would be needed to determine an accurate line list for additional electronic bands of \ch{^{32}S2} and the \ch{^{32}S^{34}S} isotopologue. The recent \textit{ab initio} line list of \citet{sarka_2023} accounts for isotopologues and continuum features, however the line list is insufficient at modeling the spectrum of \ch{S2} at high resolution and is unable to account for the large broadening effect caused by predissociation.

The $B''-X$ transition is included in our line list since the two $B$ and $B''$ states are heavily mixed \citep{10.1039/A606591K}. In particular, the mixing between the $B$ and $B''$ states occurs strongly where the $B$ $v'$=7-9 bands are located. The $B''-X$ transition is applied as a perturbation in the PGOPHER file, so no oscillator strengths are applied, and the intensities are borrowed from the $B-X$ transition. This gives rise to $B-X$ and $B''-X$ bands with high vibrational levels for both $v'$ and $v''$. Given that these states will be expected to have small populations, we have restricted the HITRAN-formatted line list for the $B-X$ tradition to $v' \leq 3$ and $v'' \leq 10$ below the dissociation limit and to $v' \leq 27$ and  $v'' \leq 2$ above the dissociation limit (see Fig.~\ref{fig:linelist}). The excluded bands can be recalculated using the PGOPHER file in the Supplementary Material.

There are no reported partition functions available for comparison in the literature. From a qualitative comparison with Figure 5 of \citet{2001JPhB...34.4183V} one can tell that the partition sum from our work, is multiple times larger at 1000~K, due to the necessary inclusion of spin-splitting in our model. On a related note, since our model includes energy levels up to $J_{\textrm{max}}=150$ and $v_{\textrm{max}}=10$ for the ground state, we advise caution at temperatures beyond 1500~K as we expect the partition sum to start to deviate from a complete partition function as temperature increases. The partition sum calculated for this work is included in the Supplementary Material.

%%%%%%%%%%%%%%%%%%%%%%%%%%%%%%%%%%%%%%%%%%%%%%%%%%

\section*{\small Conclusions}\vspace{-1.5mm}\hrule\vspace{3mm}
A HITRAN-formatted line list for \ch{S2} that covers the UV spectral range has been calculated from spectroscopic constants available in the literature \citep{10.1063/1.470810,10.1039/A606591K,10.1063/1.5029930} using the PGOPHER program \citep{10.1016/j.jqsrt.2016.04.010} and a fit to line positions of emission bands with a standard deviation of 0.130~cm$^{-1}$. The line list includes the prominent $B\,^{3}\Sigma^{-}_{u}-X\,^{3}\Sigma^{-}_{g}$ electronic transition of \ch{^{32}S2} with bands $v'$=0-27 and $v''$=0-10. The perturbing electronic transition $B''\,^{3}\Pi_{u}-X\,^{3}\Sigma^{-}_{g}$ is also included, with $v'$=0-19, $v''$=0-10.  The line list is provided as Supplementary Material in the commonly-used HITRAN format and will also be freely available online at \url{https://hitran.org/}.

Line intensities for predissociated bands have been obtained by fitting to the experimental observations of \citet{10.1063/1.5029929}. The line list has been validated through comparisons to existing experimental cross-section spectra for the predissociated region.   The predissociation of \ch{S2} for $v'\geq$10 in the $B-X$ transition requires the inclusion of predissociation line widths. Therefore, a Python program has been developed to be used in conjunction with HAPI that can apply the necessary predissociation line widths when using the HITRAN-formatted line list provided in the Supplementary Material. Currently, HAPI does not have the functionality to include predissociation line widths. However, it is planned that the program will be incorporated into future versions.  

Our \ch{S2} line list can be used for exoplanet and planetary atmospheric investigations and photochemical models. The inclusion of our \ch{S2} line list in photochemical models is expected to improve atmospheric interpretations of planetary spectra that have previously been based on estimates. In particular, interpretations of the JWST spectra of WASP-39b expect \ch{S2} to be a key molecule in the formation of \ch{SO2}. 

It should be noted that in May 2023, the Leiden database \citet{Heays2017} updated their \ch{S2} cross-sections to include a preliminary version of the data reported here \citep[see][]{Hrodmarsson2023}, but temperature independence still remains. The \ch{S2} line list reported in this work provides greater flexibility for temperature coverage.

In addition to new experiments (especially for minor isotopologues), in future works, \textit{ab initio} calculations (after validation) can help improve line lists of the sulfur dimer. The work of \citet{sarka_2023} is an important step in that direction.  In principle, \textit{ab initio} methods can be used to calculate line positions and intensities, but may still be deficient for perturbations. They can also provide information on the predissociation widths and continuum contribution \citep[see review by][]{Tennyson2023}.  

In summary, we provide a publicly available \ch{S2} line list and associated calculation tools, which can be used to simulate the emission and absorption spectrum of \ch{^{32}S2} over the 21\,700$-$41\,300~cm$^{-1}$ ($\sim$242$-$461~nm) spectral range. 

%%%%%%%%%%%%%%%%%%%%%%%%%%%%%%%%%%%%%%%%%%
\section*{\small Acknowledgments and Funding}\vspace{-1.5mm}\hrule\vspace{3mm}
{We respectfully acknowledge the late Colin M. Western, who created the PGOPHER program and developed the initial \ch{S2} spectroscopic model that formed the basis of this work. FMG's work was carried out in the framework of improving terrestrial exoplanet photochemical models supported through NASA 2021 Exoplanets Research Program grant (NNH21ZDA001N-XRP). We thank the PI (Edward Schwieterman) and the scientific PI (Sukrit Ranjan) of this grant for fruitful discussions and support. RJH and IEG's contribution was supported through the NASA PDART grant.  We thank Roger Yelle and Helgi Rafn Hr{\'o}{\v{o}}marsson for the discussions. We are grateful to Karolis Sarka and Shinkoh Nanbu for informing us about their very recent \textit{ab initio} calculations.}

%%%%%%%%%%%%%%%%%%%%%%%%%%%%%%%%%%%%%%%%%%
\section*{\small Conflicts of Interest}\vspace{-1.5mm}\hrule\vspace{3mm}
{The authors declare that they have no known competing financial interests or personal relationships that could have appeared to influence the work reported in this paper. The funders had no role in the design of the study; in the collection, analyses, or interpretation of data; in the writing of the manuscript, or in the decision to publish the results.}
%%%%%%%%%%%%%%%%%%%%%%%%%%%%%%%%%%%%%%%%%%

%%%%%%Bibliography%%%%%%%%%%%%%%%%%%
\section*{\small Bibliography\vspace{1.5mm}\hrule\vspace{3mm}}
\renewcommand{\bibsection}{}
\bibliographystyle{main.bst}
\bibliography{Main.bib}\label{bibliography}
\end{document}